\documentclass[journal=jacsat,manuscript=article]{achemso}
\usepackage{achemso}
\usepackage[version=3]{mhchem}
\usepackage[english]{babel}
\usepackage{filecontents}
\usepackage{graphicx}
\usepackage{courier}
\usepackage{amssymb,amsmath,amsthm,mathrsfs,comment,afterpage,accents}
\usepackage{mathtools}
\usepackage{braket}
\usepackage{mathrsfs}
\usepackage{courier}
\usepackage{color}
\usepackage{subdepth}
\usepackage{amsmath}
\setkeys{acs}{articletitle = true}
\usepackage{makecell}
\makeatletter
\SectionNumbersOn
\def\@dotsep{4.5}
\usepackage{cancel}
\makeatother
\usepackage{dcolumn}
\usepackage{bm}
\renewcommand\vec\mathbf
\newcommand\mat\mathbf
\newcommand\tr{\operatorname{tr}}

\newcommand{\insertnew}[1]{{\textcolor{black} {#1}}}
\newcommand{\revision}[1]{{\textcolor{black}{#1}}}
\newcommand{\revrep}[2]{{\textcolor{red}{}}{\textcolor{blue}{#2}}}

\title{
Two Single-Reference Approaches to Singlet Biradicaloid Problems:
Complex, Restricted Orbitals and Approximate Spin-Projection
Combined With Regularized Orbital-Optimized M{\o}ller-Plesset Perturbation Theory
}
\author{Joonho Lee}
\email{linusjoonho@gmail.com}
\author{Martin Head-Gordon}
\email{mhg@cchem.berkeley.edu}
\affiliation{
Department of Chemistry, University of California, Berkeley, California 94720, USA
Chemical Sciences Division, Lawrence Berkeley National Laboratory, Berkeley, California 94720, USA
}
\begin{document}
\newpage
\maketitle
\begin{abstract}
We present a comprehensive study of two single-reference approaches to singlet biradicaloids. 
These two approaches are based on the recently developed regularized orbital-optimized M{\o}ller-Plesset method ($\kappa$-OOMP2).
The first approach is to combine the Yamaguchi's approximate projection (AP) scheme and $\kappa$-OOMP2 with unrestricted (U) orbitals ($\kappa$-UOOMP2).
By capturing only {\it essential} symmetry breaking, $\kappa$-UOOMP2 can serve as a \insertnew{suitable} basis for AP. 
The second approach is $\kappa$-OOMP2 with complex, restricted (cR) orbitals ($\kappa$-cROOMP2). Though its applicability is \insertnew{more} limited due to the \insertnew{comparative rarity} of cR solutions, $\kappa$-cROOMP2 offers a simple framework for describing singlet biradicaloids with complex polarization while removing {\it artificial} spatial symmetry breaking.
We compare the scope of these two methods with numerical studies.
We show that AP+$\kappa$-UOOMP2 and $\kappa$-cROOMP2 can perform similarly well in the TS12 set, a data set that includes 12 data points for triplet-singlet gaps of several atoms and diatomic molecules with a triplet ground state. This was also found to be true for the barrier height of a reaction \insertnew{involving attack on} a cysteine ion by a singlet oxygen molecule. However, we also demonstrate that in highly symmetric systems like \ce{C30} (\ce{D_{5h}}) $\kappa$-cROOMP2 is more suitable as it conserves spatial symmetry. Lastly, we present an organic biradicaloid that does not have a $\kappa$-cROOMP2 solution in which case only AP+$\kappa$-UOOMP2 is applicable.
We recommend $\kappa$-cROOMP2 whenever complex polarization is {\it essential} and AP+$\kappa$-UOOMP2 for biradicaloids without {\it esssential} complex polarization but with {\it essential} spin-polarization.
\end{abstract}
\newpage
\section{Introduction} \label{sec:intro}
Strong correlation is usually \insertnew{associated with multiple} open-shell electrons that are antiferromagnetically coupled into a low-spin state.\cite{Handy2001,Hollett2011,Hollett2011a}
\insertnew{For instance, molecular magnets with multiple metal centers,\cite{Atanasov2015,Mayhall2015,Ungur2016,Frost2016,Lee2018a} non-innocent ligands,\cite{Ward2002,Butschke2015} metalloenzymes,\cite{Yachandra1996,Yamaguchi2010,Yson2013} and oligoacenes\cite{Bendikov2004,Lee2017,Schriber2018,Mullinax2019} exhibit strong correlation.}
Of such cases, singlet biradicaloids exhibit the simplest form of strong correlation.\cite{Salem1972,Slipchenko2002,Scheschkewitz2002,Bachler2002,Jung2003,Kamada2010,Abe2013}
As this is usually outside the scope of single-reference electronic structure methods, it is common to employ multiconfigurational methods.\cite{Roos1980, Ruedenberg1982, Szalay2012}
A brute-force approach to treat this strong correlation is complete active space self-consistent field (CASSCF) with an active space of two electrons in two orbitals (2e, 2o). However, CASSCF does not incorporate electron correlation outside the active space so subsequent dynamic correlation treatments\cite{Andersson1990,Nakano1993,Angeli2001} are necessary for quantitatively correct answers.
\insertnew{A related single reference approach is to start from the triplet single determinant (i.e., $M_S=1$) and flip a spin to access the $M_S=0$ manifold, either using configuration interaction (CI)\cite{Krylov2001a,Shao2003,Mayhall2014,Mayhall2015,Mato2018} or coupled-cluster (CC) via the equation of motion approach.\cite{Krylov2001,Slipchenko2002,Krylov2008}}

Alternatively, one could try to treat such systems using single-reference methods with the help of {\it essential} symmetry breaking. 
\revision{It is essential in the sense that 
the qualitative character of a single-determinant wavefunction is fundamentally wrong 
without essential breaking.} A majority of {\it essential} symmetry breaking is spin-restricted (R) to spin-unrestricted (U) symmetry breaking, namely spin-polarization. 
In the case of singet biradicaloids, such essential symmetry breaking can be combined with Yamaguchi's approximate spin-projection (AP) to produce spin-pure energies.\cite{Yamaguchi1988,Kitagawa2007,Saito2009,Nakano2010,Mak2011,Saito2012, Hratchian2013, Thompson2015}
The applicability of AP is \revrep{heavily}{}dependent on whether the underlying wavefunction contains only one contaminant.
It is an exact projection only if there is one single contaminant.
This sets a limit to $\langle\hat{S}^2\rangle$ of broken-symmetry $M_S=0$ solutions to be effective for AP: $0.0 \le \langle\hat{S}^2\rangle \le 2.0$.

UHF is heavily spin-contaminated for most biradicaloids. For instance, this was observed by us in the heptazethrene dimer (HZD) where broken-symmetry UHF yields $\langle\hat{S}^2\rangle=6.3$ in the cc-pVDZ basis set.\cite{Lee2018} The subsequent correlation treatment based on these UHF solutions via second-order M{\o}ller-Plesset perturbation theory (MP2) is not effective in removing such heavy spin contamination. It is possible to employ orbital optimized MP2 (OOMP2) as an attempt to produce a reference determinant with only essential symmetry breaking (i.e., $\langle\hat{S}^2\rangle \approx 1.0$). However, it is likely that OOMP2 produces a divergent solution or a restricted solution that is unphysically low in energy if not divergent.\cite{Stuck2013,Razban2017} 
As a solution to this problem, we employed regularized OOMP2 ($\kappa$-OOMP2) to treat HZD. \cite{Lee2018} 
\insertnew{In contrast to our previous $\delta$-OOMP2 (regularized with a constant level-shift), $\kappa$-OOMP2 determines the strength of regularization of
individual correlation energy contributions depending on the orbital energy gap associated with them. 
$\kappa$-OOMP2, in turn, achieves both the recovery of Coulson-Fischer points\cite{Coulson1949} and favorable thermochemistry performance, which was found to be challenging for $\delta$-OOMP2 to achieve.\cite{Razban2017}
}
Returning to the HZD example, $\kappa$-OOMP2 with unrestricted orbitals ($\kappa$-UOOMP2) produces $\langle\hat{S}^2\rangle=1.2$ which is well-suited for subsequent AP treatment. Generally speaking, $\kappa$-UOOMP2 with AP (AP+$\kappa$-UOOMP2) is a simple and robust way to treat biradicaloids which captures both static and dynamic correlation.
We will further highlight this particular combination of AP and $\kappa$-UOOMP2 later in this work.

A rather rarer class of {\it essential} symmetry breaking, which is another focus of this work, is real, R to complex, R (cR) symmetry breaking. This is referred to as ``complex-polarization'' in this work. Complex polarization was known for many years in the context of some strongly correlated \insertnew{molecules such as  \ce{O2} ($^1\Delta_g$).\cite{Radom1973,Fukutome1973,Haddon1975,Dill1975,Bohm1981,Bohm1981a,Bohm1983,Krogh-Jespersen1985,Mains1990}} Our group established its connection to generalized valence bond perfect pairing (GVB-PP)\cite{Goddard1973} using the complex pairing theorem.\cite{Small2015a} When such solutions exist, complex restricted Hartree-Fock (cRHF) can indeed capture some aspects of GVB-PP and behaves qualitatively better than RHF. It was shown that the subsequent correlation treatment, cRMP2, yields quantitatively more accurate results than RMP2 for systems examined in ref. \citenum{Small2015a}. Moreover, cRMP2 outperformed UMP2 especially when there is a strong mixing between singlet and triplet states. 

Our recent work illustrated a way to obtain such essential symmetry breaking with $\kappa$-OOMP2.\cite{Lee2019} Therein we discussed how to remove {\it artificial} spin-polarization using $\kappa$-OOMP2 with complex, generalized (cG) orbitals. 
\revision{It is artificial because orbital optimization in the presence of dynamic correlation such as MP2 (or other approaches for approximate Br{\"u}ckner orbitals) may remove such symmetry breaking.
Artificial symmetry breaking occurs at the HF level not due to the lack of ability to describe strong correlation but because of the lack of dynamic correlation treatment. In ref. \citenum{Lee2019}, we show that it is possible to distinguish artificial and essential symmetry breaking based on $\kappa$-OOMP2. Interested readers are referred to ref. \citenum{Lee2019} and we will further review some aspects of this relevant to this work in Section \ref{sec:oomp2}.} 
\revision{In addition to {\it essential} spin-symmetry breaking,} it is also possible to explore {\it essential} complex-polarization within the $\kappa$-OOMP2 method, which will combine the strengths of cRMP2 and $\kappa$-OOMP2. Namely, $\kappa$-cROOMP2 is able to describe multireference systems whenever complex-polarization is relevant. 

For general biradicaloid systems, it is natural to consider AP and cR methods as simple single-reference alternatives to multi-reference and spin-flip methods. In particular, these are far simpler to implement than typical multi-reference second-order perturbation theory.\cite{Andersson1990,Angeli2001} 
Compared to AP, cR methods offer more straightforward formalisms for response theory. For example, cRMP2 has the identical response theory formalism to that of usual MP2 and there is no need to derive additional terms. The analytic nuclear derivatives of AP methods have been derived and implemented at the mean-field level,\cite{Kitagawa2007,Saito2009,Saito2012, Hratchian2013} but there has been no study on response theory of correlated wavefunction methods with AP. While the formal and practical simplicity of cR methods is very desirable, its limited applicability due to the rareness of cR solutions makes it less appealing. 

In this work we will explore several biradicaloid systems that exhibit cRHF solutions and discuss the applicability of $\kappa$-cROOMP2 and AP+$\kappa$-UOOMP2. In particular, we will compare $\kappa$-cROOMP2 and AP+$\kappa$-UOOMP2 in these systems and discuss the similarities and differences between them. For simplicity, we will limit our discussion to HF, MP2 and $\kappa$-OOMP2 although other variants of MP2 and OOMP2, such as spin-component scaled methods,\cite{Grimme2003, Jung2004} can also be combined with cR orbitals or AP.

%

\section{Theory}
We will use $i,j,k,l,\cdot\cdot\cdot$ to index occupied orbitals, $a,b,c,d,\cdot\cdot\cdot$ to index virtual orbitals, and $p,q,r,s,\cdot\cdot\cdot$ to index either of those two.
\subsection{Review of cRHF and Complex Polarization}
The complex restricted Hartree-Fock (cRHF) energy is given by
\begin{equation}
E_\text{HF} = 2 \tr\left(\mathbf P \mathbf H_0\right) + 2 \tr \left( \mathbf P \mathbf J\right) - \tr \left( \mathbf P\mathbf K\right) + E_\text{nuc}
\label{eq:ehf}
\end{equation}
where $\mathbf P$ is the one-particle reduced density matrix (1PDM), $\mathbf H_0$ is the one-electron Hamiltonian, $\mathbf J$ and $\mathbf K$ are the Coulomb and exchange matrices, and $E_\text{nuc}$ is the nuclear repulsion energy.
In cRHF, we allow the molecular orbital (MO) coefficient matrix $\mathbf C$ to be complex and as a result $\mathbf P$ may become complex.
As mentioned in ref. \citenum{Small2015a}, a cRHF solution is ``fundamentally complex'' if and only if the norm of the imaginary part of $\mathbf P$ is non-zero.

The use of complex restricted (cR) orbitals for multi-reference problems has been known for many years in electronic structure theory\cite{Radom1973,Fukutome1973,Haddon1975,Dill1975,Bohm1981,Bohm1981a,Bohm1983,Krogh-Jespersen1985,Mains1990} but they have been rarely employed in practice. The major reason for this underappreciation is due to the rareness of genuine cR solutions.
Small et al. established the connection between cRHF and GVB-PP and as a result, we have a better understanding of why cR solutions are rare and when to expect them.\cite{Small2015a}

Within a single pair of electrons, the R to cR instability is driven by the energy lowering due to a PP-like (or CAS(2,2)-like) configuration. 
However, a cRHF wavefunction also necessarily contains an open-shell singlet (OSS)-like configuration which is usually energetically high. 
The competition between the PP-like contribution \insertnew{(energy-lowering versus R)} and the OSS-like contribution \insertnew{(energy-raising versus R)} determines the R to cR instability. 
When the PP stabilization is greater than the OSS energetic cost, we observe the R to cR instability. 
This is, however, not very common to observe and this explains the rareness of cRHF solutions.
As we will see, some singlet biradicaloids exhibit complex-polarization and therefore cRHF can serve as a faithful starting point for subsequent correlation treatments.
The relative energetics between PP-like terms and OSS-like terms change in the presence of correlation treatment. 
Therefore, it is reasonable to expect that some cRHF solutions are {\it artificial} and they would lead to cRMP2 energies that are much higher than RMP2.
We will encounter an example that demonstrates this later in the paper.

It is useful to run internal stability analysis to ensure the local stability of cRHF solutions. We provide the electronic Hessian of the energy expression in Eq. \eqref{eq:ehf} in Appendix.
\subsection{Regularized OOMP2 with cR orbitals: $\kappa$-cROOMP2}\label{sec:oomp2}
The MP2 energy expression with cRHF orbitals reads
\begin{equation}
E_\text{cRMP2} =  E_\text{HF} + \sum_{ijab} {\tau_{ij}^{ab}} \left(ia|jb\right)
\label{eq:emp2}
\end{equation}
where $i$ and $j$ are occupied spatial orbitals, $a$ and $b$ are unoccupied spatial orbitals, $\left(ia|jb\right)$ represents the two-electron four-center integrals and the spin-adapted amplitudes $\tau$ are
\begin{equation}
\tau_{ij}^{ab} =- \frac{2\left(ia|jb\right)^* - \left(ib|ja\right)^*}{\Delta_{ij}^{ab}}.
\end{equation}
$\Delta_{ij}^{ab}$ is a positive energy denominator defined as
\begin{equation}
\Delta_{ij}^{ab} = \epsilon_a + \epsilon_b - \epsilon_i - \epsilon_j,
\end{equation}
where $\epsilon_p$ denotes \insertnew{canonical} orbital energies.
Orbital optimization of Eq. \eqref{eq:emp2} yields orbital-optimized MP2 (OOMP2). 
As mentioned in Section \ref{sec:intro}, OOMP2 has two major issues that limits its applicability. First, as we optimize orbitals in the presence of correlation energy, $\Delta_{ij}^{ab}$ can become very small and the resulting energy \insertnew{can become non-variational and even approach divergence}.\cite{Stuck2013} Second, as a result OOMP2 may {\it unphysically} prefer restricted solutions and remove the Coulson-Fischer point.\cite{Razban2017} Our group has developed a regularization scheme which fixes these two major issues in OOMP2.\cite{Lee2018}

The \insertnew{orbital-energy-dependent} regularization introduced in ref. \citenum{Lee2018} modifies the two-electron integrals in the correlation energy contribution in Eq. \eqref{eq:emp2}. The resulting $\kappa$-cRMP2 energy expression reads
\begin{equation}
E_\text{$\kappa$-cRMP2} =  E_{HF} + \sum_{ijab} {\tilde{\tau}_{ij}^{ab}}^* \left(ia|jb\right)(1 - e^{-\kappa\Delta_{ij}^{ab}})
\label{eq:kemp2}
\end{equation}
where the regularized amplitudes are
\begin{equation}
\tilde{\tau}_{ij}^{ab} = \tau_{ij}^{ab}(1 - e^{-\kappa\Delta_{ij}^{ab}})
\end{equation}
Orbital optimizing Eq. \eqref{eq:kemp2} defines the $\kappa$-OOMP2 method (in this case $\kappa$-cROOMP2).
It is immediately obvious that the correlation energy can no longer diverge even when $\Delta_{ij}^{ab} = 0$. 
Based on carbon-carbon single, double, and triple bond breaking, we showed that the Coulson-Fischer point is recovered.
\insertnew{Combining recovery of Coulson-Fischer points with reasonable performance for a} thermochemistry benchmark, $\kappa=1.45$ was recommended for chemical applications.\cite{Lee2018} We also showed that \revision{$\kappa \in [1.0, 2.0]$ (which comfortably includes $\kappa=1.45$ in the middle)} yields only {\it essential} symmetry breaking and can remove {\it artificial} HF symmetry breaking in fullerenes.\cite{Lee2019} \insertnew{This is because $\kappa$-OOMP2 describes dynamic correlation, but regularization has removed the inaccurate description of static correlation present in conventional MP2.}

\revision{Distinguishing artificial and essential symmetry breaking based on $\kappa$-OOMP2 may seem arbitrary. However, in ref. \citenum{Lee2019}
we compared this diagnosis of strong correlation with other approaches such as natural orbital occupation numbers and more sophisticated coupled-cluster methods. All these three independent probes suggested that \ce{C60} is not strongly correlated and \ce{C36} is strongly correlated.
As such, $\kappa$-OOMP2 can reliably probe the underlying symmetry breaking and answer whether it is artificial (not strongly correlated) or essential (strongly correlated). $\kappa$-OOMP2 will be used to probe essential symmetry breaking and strong correlation in another fullerene \ce{C30} below.}

The implementation of $\kappa$-cROOMP2 was accomplished closely following the spin-orbital implementation described in ref. \citenum{Lee2018}. We apply the resolution-of-the-identity approximation to $\left(ia|jb\right)$,
\begin{equation}
\left(ia|jb\right) =\sum_{PQ}\left(ia|P\right) (P|Q)^{-1}\left(Q|jb\right) = \sum_P (ia|P) C_{jb}^{P}
\end{equation}
where $P$ and $Q$ are auxiliary basis indices and we define \insertnew{the expansion coefficients of an occupied-virtual product $|jb)$ as:}
\begin{equation}
C_{jb}^{P} = \sum_Q (P|Q)^{-1}\left(Q|jb\right)
\end{equation}
The spin-adapted two-particle density matrix (2PDM) consists of two parts:
one is the usual MP2 2PDM contribution,
\begin{equation}
\Gamma_{ai}^{P} = 2\sum_{jb}C^{P}_{jb}\tilde{\tau}_{ij}^{ab},
\label{eq:gamma1}
\end{equation}
and another is the modification due to the regularizer,
\begin{equation}
\tilde{\Gamma}_{ai}^{P} = 
2\sum_{jb} C^{P}_{jb}\tilde{\tau}_{ij}^{ab} e^{-\kappa(\epsilon_b - \epsilon_j)}
\label{eq:gamma2}
\end{equation}
Similarly, the spin-adapted 1PDM also consists of two parts:
the first is the usual MP2 1PDM contributions,
\begin{equation}
P^{(2)}_{ij} = -2 \sum_{abk} ({\tilde{\tau}_{ik}^{ab}})^* \frac{(ka|jb)^*}{\Delta_{ij}^{ab}}
\label{eq:poo}
\end{equation}
\begin{equation}
P^{(2)}_{ca} = 2 \sum_{ijb} ({\tilde{\tau}_{ij}^{ab}})^* \frac{(ic|jb)^*}{\Delta_{ij}^{cb}}
\label{eq:pvv}
\end{equation}
and the second is the modification from the regularizer,
\begin{align}\label{eq:magic1}
\tilde{P}^{(2)}_{ij} &= -\kappa\int_0^1 \text{d}\tau e^{\tau\kappa\epsilon_i} (\omega_{ij}^* + \omega_{ji}) e^{(1-\tau)\kappa\epsilon_j} 
= - (\omega_{ij}^* + \omega_{ji}) \left(\frac{e^{\kappa\epsilon_i} - e^{\kappa\epsilon_j}}{\epsilon_i-\epsilon_j}\right)
\\
\label{eq:magic2}
\tilde{P}^{(2)}_{ab} &= \kappa\int_0^1 \text{d}\tau e^{-\tau\kappa\epsilon_a} (\omega_{ab} + \omega_{ba}^*) e^{-(1-\tau)\kappa\epsilon_b}
= (\omega_{ab} + \omega_{ba}^*) \left(\frac{e^{-\kappa\epsilon_a} - e^{-\kappa\epsilon_b}}{\epsilon_b-\epsilon_a}\right)
\end{align}
where the definition of $\omega_{ij}$ and $\omega_{ab}$ follows: 
\begin{equation}
\omega_{ij} = \sum_{aP}e^{-\kappa\epsilon_a} (ia|P)
\tilde{\Gamma}^P_{aj}
\end{equation}
and
\begin{equation}
\omega_{ab} = \sum_{iP}e^{\kappa\epsilon_i} \tilde{\Gamma}^P_{ai}(ib|P)
\end{equation}
These spin-adapted quantities can be used to produce appropriate orbital gradients for orbital optimization. 
Interested readers are referred to ref. \citenum{Lee2018} for more technical details.
In passing we mention that Eq. \eqref{eq:magic1} and Eq. \eqref{eq:magic2} were computed via a one-dimensional Legendre quadrature previously,\cite{Lee2018} but in the pseudocanonical basis,
it can be done analytically as shown above.

\insertnew{We apply the frozen-core approximation to the systems considered in this paper. 
This adds orbital rotation parameters between frozen core and occupied orbitals to the orbital optimization problem.
We present the pertinent orbital gradient equations and explain some numerical difficulties we encountered with this optimization problem in the Supporting Information.}

\subsection{Yamaguchi's Approximate Spin-Projection}
The approximate spin-projection method proposed by Yamaguchi \cite{Yamaguchi1988} has been widely used in a wide variety of strong correlation problems.\cite{Yamaguchi1988,Kitagawa2007,Saito2009,Nakano2010,Mak2011,Saito2012, Hratchian2013, Thompson2015}
Its working equation is very simple and it usually takes at most two separate single point calculations for two different $M_S$ values to perform the projection.
When projecting a triplet state out of an $M_S=0$ broken symmetry solution, one can use the following equation which is derived using $\langle S^2 \rangle$:
\begin{equation}
E_{S=0}  = \frac{E_\text{BS} - (1- \alpha) E_{S=1}}{\alpha}
\label{eq:Eproj}
\end{equation}
where the spin-coupling coefficient $\alpha$ is
\begin{equation}
\alpha = \frac{\langle S^2\rangle_{S=1} - \langle S^2 \rangle _\text{BS}}{\langle S^2\rangle_{S=1} - \langle S^2\rangle_{S=0}}
\end{equation}
There are multiple ways to obtain $E_{S=1}$ and $\langle S^2 \rangle_{S=1}$. The simplest way is to 
\insertnew{use a high spin $M_S=1$ calculation to obtain
$E_{S=1}$ at the same level of theory as $E_\text{BS}$.}
Therefore, we need a total of two unrestricted calculations, $M_S=0$ and $M_S=1$.
Evidently, if the singlet is heavily spin-contaminated the above spin-coupling equation is no longer valid.
Furthermore, we need a nearly spin-pure value of $\langle S^2 \rangle$ for the $M_S=1$ state.
As we shall see later, $\kappa$-UOOMP2 can accomplish these objectives.
\section{Applications}
We will study multiple biradicaloid systems that have one pair of electrons that exhibit {\it essential} complex-polarization or spin-polarization.
In other words, the singlet ground state of these systems involve a pair of open-shell electrons.
Throughout the examples presented below, we will see how $\kappa$-cROOMP2 and/or AP+$\kappa$-UOOMP2 can be used for these singlet biradicaloids and also compare their strengths and weaknesses.

All calculations were performed with a development version of Q-Chem.\cite{Shao2015} For $\kappa$-OOMP2 methods, we took a stable HF solution as an initial set of orbitals unless mentioned otherwise. 
 All plots were generated with $\texttt{Matplotlib}$ \cite{Hunter2007} and all molecular figures were generated with $\texttt{Chemcraft}$.\cite{Chemcraft} 
\subsection{TS12 Set: Triplet-Singlet Gaps}\label{sec:st12}

We will consider triplet-singlet gaps ($\Delta E_\text{T-S} = E_\text{S} - E_\text{T}$) of atoms and diatomics whose ground state is triplet. Systems with a triplet ground state are likely to have a (near) degeneracy between highest occupied molecular orbital (HOMO) and lowest unoccupied molecular orbital (LUMO) so these are also likely to have near-degenerate OSS-like and PP-like configurations. Therefore, for these molecules, there is a good chance for {\it essential} complex polarization to occur.

We will compare HF, MP2, and $\kappa$-OOMP2 methods with different types of orbitals for treating the singlet ground state of the following molecules: \ce{C}, \ce{NF}, \ce{NH}, \ce{NO-}, \ce{O2}, \ce{O}, \ce{PF}, \ce{PH}, \ce{S2}, \ce{S}, \ce{Si}, and \ce{SO}. The reference triplet-singlet gaps as well as the equilibrium bond length of diatomics for each electronic state were taken from the NIST Chemistry WebBook.\cite{nist}
Individual references for these experimental values and geometries are given in the Supporting Information.
We validated the experimental gaps against near-exact full configuration interaction calculations using the heat-bath algorithm developed by Holmes and co-workers.\cite{Holmes2016} The theoretical estimation lies within 1.0 kcal/mol of the experimental values and we provide these data in the Supporting Information. 
This data set will be referred to as the ``TS12'' set for the rest of this manuscript.

In benchmarking HF, MP2, and $\kappa$-OOMP2 methods, we employed the aug-cc-pVQZ basis set\cite{Dunning1989, Kendall1992} along with its auxiliary basis set.\cite{Weigend2002} The frozen core approximation was used for all correlated wavefunction calculations. Unrestricted orbitals are used for the triplet state ($M_S=1$).

\begin{table}[h!]
  \centering
  \begin{tabular}{c|r|r|c|c}\hline
 & $M_S=0$ & $M_S=1$ & Triplet & Singlet\\\hline
\ce{C} & 1.018 & 2.010 & ${}^3$P & ${}^1$D\\\hline
\ce{NF} & 1.015 & 2.023 & X${}^3\Sigma^-$ & a${}^1\Delta$\\\hline
\ce{NH} & 1.012 & 2.017 & X${}^3\Sigma^-$ & a${}^1\Delta$\\\hline
\ce{NO-} & 1.031 & 2.052 & X${}^3\Sigma^-$ & a${}^1\Delta$\\\hline
\ce{O2} & 1.023 & 2.049 & X${}^3\Sigma^-_\text{g}$ & a${}^1\Delta_\text{g}$\\\hline
\ce{O} & 1.009 & 2.009 & ${}^3$P & ${}^1$D\\\hline
\ce{PF} & 1.047 & 2.035 & X${}^3\Sigma^-$ & a${}^1\Delta$\\\hline
\ce{PH} & 1.039 & 2.029 & X${}^3\Sigma^-$ & a${}^1\Delta$\\\hline
\ce{S2} & 1.062 & 2.060 & X${}^3\Sigma^-_\text{g}$ & a${}^1\Delta_\text{g}$\\\hline
\ce{S} & 1.033 & 2.013 & ${}^3$P & ${}^1$D\\\hline
\ce{Si} & 1.047 & 2.015 & ${}^3$P & ${}^1$D\\\hline
\ce{SO} & 1.051 & 2.058 & X${}^3\Sigma^-$ & a${}^1\Delta$\\\hline
  \end{tabular}
  \caption{
The UHF $\langle S^2 \rangle$ values of the molecules in the test set considered in this work and the term symbol for each electronic state considered in the TS12 set.}
  \label{tab:s2}
\end{table}
For the molecules in the TS12 set, using real, restricted orbitals for the singlet ground state is fundamentally incorrect as it cannot capture the biradicaloid character of the singlet ground state. UHF orbitals are heavily spin-contaminated as the singlet ground state is a strong biradicaloid. This is well illustrated in Table \ref{tab:s2}. The $M_S=0$ states exhibit $\langle S^2 \rangle = 1.0$ which indicates nearly perfect singlet biradicals. The $M_S=1$ states are more or less spin-pure which validates the use of UHF orbitals for $M_S=1$ states.
Therefore, UHF and UMP2 are expected to perform very poorly on this test set.
However, all these $\langle S^2 \rangle$ values are very well-suited for the AP approach. Therefore, one may expect that AP+UMP2 and AP+$\kappa$-UOOMP2 perform similarly well. We will see whether these predictions are indeed true in the TS12 set.

\begin{table}[h!]
  \centering
  \begin{tabular}{c|r|r|r|r|r}
   & Expt. & RHF & UHF & RMP2 & UMP2 \\\Xhline{3\arrayrulewidth}
\ce{C} & 29.14 & 26.59 & -15.37 & 13.85 & -13.58 \\\hline
\ce{NF} & 34.32 & 31.54 & -14.80 & 10.99 & -17.23 \\\hline
\ce{NH} & 35.93 & 30.59 & -16.72 & 15.90 & -17.29 \\\hline
\ce{NO-} & 17.30 & 29.60 & -2.11 & 5.53 & -7.74 \\\hline
\ce{O2} & 22.64 & 32.54 & -5.45 & 6.15 & 2.72 \\\hline
\ce{O} & 45.37 & 34.72 & -22.79 & 19.71 & -22.10 \\\hline
\ce{PF} & 20.27 & 25.37 & -11.89 & 10.80 & -9.06 \\\hline
\ce{PH} & 21.90 & 24.35 & -11.93 & 11.66 & -10.17 \\\hline
\ce{S2} & 13.44 & 21.03 & -5.70 & 4.48 & -5.01 \\\hline
\ce{S} & 26.41 & 26.52 & -15.75 & 14.21 & -12.19 \\\hline
\ce{Si} & 18.01 & 20.13 & -11.77 & 10.12 & -7.76 \\\hline
\ce{SO} & 18.16 & 24.77 & -6.94 & 3.94 & -9.84 \\\Xhline{3\arrayrulewidth}
RMSD & N/A & 27.66 & 13.04 & 11.60 & 12.42 \\\hline
MSD & N/A & 27.31 & -11.77 & 10.61 & -10.77 
  \end{tabular}
  \caption{
The experimental triplet-singlet gap $\Delta E_\text{T-S} (= E_S - E_T)$ (kcal/mol) of various atoms and diatomics and the deviation (kcal/mol) in 
$\Delta E_\text{T-S}$ obtained with HF and MP2 using restricted and unrestricted orbitals. RMSD stands for root-mean-square-deviation and MSD stands for mean-signed-deviation.}
  \label{tab:st1}
\end{table}

First, we discuss HF and MP2 with real, restricted (R) and real, unrestricted orbitals (U). The results of these methods are presented in Table \ref{tab:st1}. Based on the mean-signed-deviation (MSD) of each method, it is evident that restricted orbitals overestimate the gap whereas unrestricted orbitals underestimate the gap. This suggests that the singlet ground state of these molecules is too high in energy when described by R orbitals and too low in energy when described by U orbitals. This is expected for RHF because closed-shell electronic structure produced by R orbitals should be less stable than an open-shell one. It is also expected for UHF, as the triplet ground state is lower in energy than the singlet ground state, triplet-singlet spin contamination lowers the energy of $M_S=0$ unrestricted state. 
With the MP2 level of correlation, these failures of R and U orbitals do not disappear. RMP2 has an RMSD of 11.60 kcal/mol and UMP2 has an RMSD of 12.42 kcal/mol. 

\begin{table}[h!]
  \centering
  \begin{tabular}{c|r|r}
 & $\kappa$-ROOMP2 & $\kappa$-UOOMP2\\\Xhline{3\arrayrulewidth}
\ce{C} & 15.71 & -13.97\\ \hline
\ce{NF} & 12.07 & -17.31\\ \hline
\ce{NH} & 17.46 & -17.04\\ \hline
\ce{NO-} & 10.59 & -6.71\\ \hline
\ce{O2} & 11.18 & -10.18\\ \hline
\ce{O} & 20.67 & -22.02\\ \hline
\ce{PF} & 14.06 & -9.85\\ \hline
\ce{PH} & 14.68 & -10.53\\ \hline
\ce{S2} & 10.34 & -5.25\\ \hline
\ce{S} & 16.19 & -12.86\\ \hline
\ce{Si} & 12.91 & -9.43\\ \hline
\ce{SO} & 10.09 & -8.54\\\Xhline{3\arrayrulewidth}
RMSD & 14.18 & 12.85\\ \hline
MSD & 13.83 & -11.97
  \end{tabular}
  \caption{
The deviation (kcal/mol) in $\Delta E_\text{T-S} (= E_S - E_T)$ obtained with different MP2 and OOMP2 methods with complex, restricted (cR) orbitals. RMSD stands for root-mean-square-deviation and MSD stands for mean-signed-deviation.}
  \label{tab:st2}
\end{table}

How does $\kappa$-OOMP2 change this conclusion? As long as R or U orbitals are employed, very similar behavior is observed. 
As it is typical for RMP2 to overestimate correlation energies for singlet biradicaloids, we expect $\kappa$-ROOMP2 to produce larger triplet-singlet gaps than those of RMP2. This is mainly due to the regularization which is more effective on the singlet states here. 
Since none of the systems exhibit {\it artificial} spin-symmetry breaking (as presented in Table \ref{tab:s2}), it is expected that $\kappa$-UOOMP2 methods do not significantly change the energetics of these systems.

In Table \ref{tab:st2}, we see that the $\kappa$-ROOMP2 gaps are all greater than the RMP2 gaps in Table \ref{tab:st1}, which confirms our prediction.
For $\kappa$-UOOMP2, the gaps are all within 2 kcal/mol from those of UMP2 except \ce{O2}. In \ce{O2}, the difference between these two methods is 12.90 kcal/mol. 
This is due to the underlying {\it artificial} reflection spatial symmetry breaking in addition to the {\it essential} spin symmetry breaking in the UHF $M_S=0$ solution.
The {\it artificial} symmetry breaking is removed with $\kappa$-UOOMP2 while the essential one still persists. 

\begin{table}[h!]
  \centering
  \begin{tabular}{c|r|r|r|r|r|r}
 & cRHF & AP+UHF & cRMP2 & AP+UMP2 & $\kappa$-cROOMP2 & AP+$\kappa$-UOOMP2\\ \Xhline{3\arrayrulewidth}
\ce{C} & 9.83 & -1.24 & 1.36 & 3.61 & 2.04 & 2.38\\ \hline
\ce{NF} & 12.71 & 4.86 & -1.70 & 1.41 & -1.28 & 0.93\\ \hline
\ce{NH} & 11.04 & 2.63 & 0.59 & 3.14 & 1.44 & 3.38\\ \hline
\ce{NO-} & 17.42 & 13.21 & -0.72 & 2.50 & 2.74 & 4.41\\ \hline
\ce{O2} & 17.85 & 11.69 & -2.26 & 29.34 & 1.50 & 3.02\\ \hline
\ce{O} & 10.44 & -0.01 & 0.65 & 3.51 & 1.04 & 3.38\\ \hline
\ce{PF} & 12.62 & -3.01 & 0.94 & 3.49 & 3.42 & 1.38\\ \hline
\ce{PH} & 11.41 & -1.45 & 0.91 & 2.98 & 3.26 & 1.73\\ \hline
\ce{S2} & 12.59 & 2.53 & -1.70 & 4.22 & 3.22 & 3.33\\ \hline
\ce{S} & 11.22 & -4.53 & 1.43 & 3.79 & 2.73 & 1.81\\ \hline
\ce{Si} & 9.10 & -5.03 & 1.45 & 3.86 & 3.27 & -0.18\\ \hline
\ce{SO} & 13.89 & 4.76 & -3.49 & -0.79 & 1.50 & 1.63\\ \Xhline{3\arrayrulewidth}
RMSD & 12.78 & 5.98 & 1.64 & 9.00 & 2.45 & 2.58\\ \hline
MSD & 12.51 & 2.03 & -0.21 & 5.09 & 2.07 & 2.27
  \end{tabular}
  \caption{
The deviation (kcal/mol) in $\Delta E_\text{T-S} (= E_S - E_T)$ obtained with HF, MP2, and $\kappa$-OOMP2 with approximate spin-projection (AP) and complex, restricted (cR) orbitals. Note that the AP procedure was carried out using the first-order corrected spin expectation values in the case of UMP2 and $\kappa$-UOOMP2. 
RMSD stands for root-mean-square-deviation and MSD stands for mean-signed-deviation. 
}
  \label{tab:st3}
\end{table}
We discuss whether these unrestricted states serve as reasonable bases to apply AP as well as whether cR orbitals can improve these catastrophic failures of HF, MP2, and $\kappa$-OOMP2 with R and U orbitals.
The results of cR and AP methods are presented in Table \ref{tab:st3}.
Neither cRHF nor AP+UHF produces satisfying results due to the lack of dynamic correlation.
Moreover, cRHF and AP+UHF show significant differences in all molecules (the smallest difference is 4.21 kcal/mol and the largest one is 15.75 kcal/mol!).
With MP2, cRMP2 is quite satisfying in that it has an RMSD of 1.64 kcal/mol with an MSD of -0.21 kcal/mol. The TS12 set can indeed be described properly with cR orbitals. On the other hand, the performance of AP+UMP2 is somewhat disappointing as it is poorer than cRMP2. In particular, an error of 29.34 kcal/mol in the case of \ce{O2} is a striking outlier. This is due to spatial symmetry breaking in UHF $M_S=0$ which cannot be fixed by UMP2 but can be fixed by $\kappa$-UOOMP2. Other than \ce{O2}, we observe a non-negligible difference (5.92 kcal/mol) in \ce{S2} which is also caused by spatial symmetry breaking in the UHF solution. All the other molecules exhibit 2-3 kcal/mol differences between these two methods.

Orbital optimization in the presence of MP2 yields significantly better AP results but $\kappa$-cROOMP2 produces slightly worse results than cRMP2.
The slight degradation in performance of cRMP2 in $\kappa$-cROOMP2 shows an interesting trend. All data points show larger triplet-singlet gaps with $\kappa$-cROOMP2 than with cRMP2. This indicates that there may be some overcorrelation problems with cRMP2 which is being regularized by $\kappa$-cROOMP2. 
Given the substantially better performance of $\kappa$-cROOMP2 compared to its R and U versions, this result is still very encouraging. Moreover, we emphasize that it is only $\kappa$-OOMP2 orbitals that yield quantitatively similar results between cR and AP approaches by harnessing only {\it essential} symmetry breaking.

For the rest of this work, we will further numerically show the quantitative similarity between AP+$\kappa$-UOOMP2 and $\kappa$-cROOMP2 beyond model systems.

\subsection{Reactivity of Deprotonated Cysteine Ion with \ce{O2} ($^1\Delta_g$)}
There are not so many chemical systems for which cR methods can be a useful alternative to standard multi-reference methods.
Any systems involving singlet oxygen (\ce{O2} ($^1\Delta_g$)) are good candidates.
In particular, singlet oxygen appears frequently in reactions in biological systems. \cite{Ogilby2010}
An example that we will study here is the reaction between an amino acid, cysteine (Cys) and singlet oxygen.
Cys is one of the five amino acids that are susceptible to singlet oxygen attack.\cite{Davies2004}
Because of the multi-reference nature of singlet oxygen, studying reactivity of Cys is challenging for single-reference methods.
As shown in Section \ref{sec:st12}, \ce{O2} ($^1\Delta_g$) exhibits {\it essential} complex polarization.
Therefore, this is an interesting example for comparing AP and cR approaches.

Lu et al. studied the reactivity of {Cys ions} with \ce{O2} ($^1\Delta_g$) using Yamaguchi's AP.\cite{Lu2017} 
As mentioned earlier, In the case of singlet oxygen, the only spin contaminant is the triplet ground state. Therefore, AP is well-suited for this case. What Lu and co-workers found is that the reactivity of Cys ions with singlet oxygen is much smaller than that of neutral Cys. This was shown by a high activation barrier along a reactive pathway.

\begin{figure}[h!]
\includegraphics[scale=0.5]{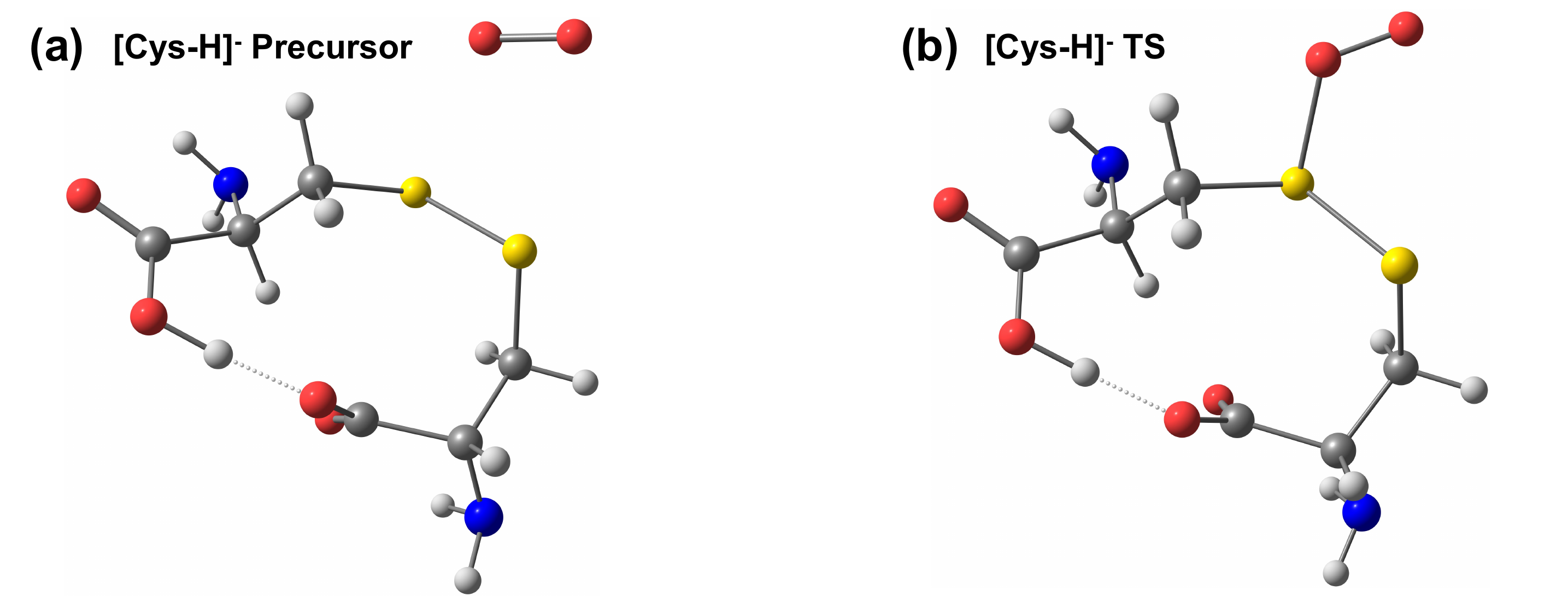}
\caption{
Molecular geometries for (a) the precursor and (b) the transition state (TS) for [Cys-H]${}^{-}$ + \ce{O2}. The Cartesian coordinates for each geometry were taken from ref. \citenum{Lu2017}.
}
\label{fig:small1}
\end{figure}

We will study a reaction between deprotonated Cys ([Cys-H]${}^{-}$) and singlet oxygen. Although there are multiple local minima geometries available, we investigated the lowest energy geometries from among those which Lu and co-workers reported. The molecular geometries of the precursor and transition state are shown in Figure \ref{fig:small1}. 
Lu and co-workers optimized the geometries at the level of B3LYP with the 6-31+G(d) basis set {\it with restricted orbitals}. 

The precursor in Figure \ref{fig:small1} has substantial open-shell character due to the presence of singlet oxygen, but the transition state (TS) is a closed-shell molecule because of the formation of a persulfoxide. It is possible that the geometry optimization of the precursor (Figure \ref{fig:small1} (a)) may produce a qualitatively wrong geometry when performed with restricted orbitals.
We independently investigated this using unrestricted orbitals and could not find a local minimum similar to Figure \ref{fig:small1} (a).
A precise determination of the precursor geometry would be interesting to study in the future using cR orbitals or AP methods.

Nonetheless, for present purposes we studied this system using the RB3LYP geometries from those of Lu and co-workers \insertnew{for single point cR and AP calculations}.
We employed the cc-pVTZ basis set\cite{Dunning1989} and the associated auxiliary basis set.\cite{Weigend2002} For the computational efficiency, the frozen core approximation was used for correlated wavefunction calculations.
The goal of our study is to demonstrate the power of cR orbitals in comparison to AP methods \insertnew{(and conventional R and U orbitals) for the open-shell singlet precursor geometry.}

\begin{table}[h!]
  \centering
  \begin{tabular}{c|r|r}\Xhline{3\arrayrulewidth}
  Method & $\Delta E$ & $\langle \hat{S}^2\rangle_{M_S=0}$\\\Xhline{3\arrayrulewidth}
RHF & 9.79 & 0.000\\\hline
cRHF & 16.93 & 0.000\\\hline
UHF & 45.17 & 1.023\\\hline
AP+UHF & 33.71 & \\\hline
  \end{tabular}
  \caption{
The activation energy $\Delta E$ (kcal/mol) of [Cys-H]$^{-}$+\ce{O2} from various types of HF.
The expectation values of $\langle \hat{S}^2\rangle$ for the $M_S=0$ state of the precursor are presented as well. 
  }
  \label{tab:hfsinglet}
\end{table}

We first discuss how different types of HF methods perform in predicting the reaction energy barrier (i.e., $E$(TS) - $E$(precursor)). We compare the use of R, U, and cR orbitals for the precursor. The precursor RHF energy should be much higher than cRHF whereas the UHF energy should be too low since 
the triplet contaminant is much more stable. Therefore, a back-of-the-envelope estimation for the relative energy barrier ordering is RHF $<$ cRHF $<$ UHF. This is indeed supported by numerical results presented in Table \ref{tab:hfsinglet}.

The relative activation energy ordering will change based on the subsequent correlation treatment.
For instance, UHF orbitals are heavily spin-contaminated so the subsequent UMP2 correlation energy will be underestimated, which then leads to a substantially smaller energy barrier than for UHF.
Similarly, RHF orbitals should also lead to somewhat high energy when combined with MP2, which then yields a smaller energy barrier than that of cRMP2.
Therefore, it is expected that the relative energy barrier ordering of MP2 methods is cR $>$ R $>$ U.

\begin{table}[h!]
  \centering
  \begin{tabular}{c|r|r}\Xhline{3\arrayrulewidth}
  Method & $\Delta E$ & $\langle \hat{S}^2\rangle_{M_S=0}$\\\Xhline{3\arrayrulewidth}
RMP2 & 19.89 & 0.000\\\hline
cRMP2 & 19.47 & 0.000\\\hline
UMP2 & 4.04 & 1.024\\\hline
AP+UMP2 & -18.32 & \\\hline
  \end{tabular}
  \caption{
The activation energy $\Delta E$ (kcal/mol) of [Cys-H]$^{-}$+\ce{O2} from various types of HF.
The expectation values of $\langle \hat{S}^2\rangle$ for the $M_S=0$ state of the precursor are presented as well. 
  }
  \label{tab:mp2singlet}
\end{table}

In Table \ref{tab:mp2singlet}, we observe the following trend instead: cR $\approx$ R $>$ U. 
It is perhaps surprising that cR and R produce more or less the same energy barriers. 
There is quite strong complex polarization within a pair of electrons which led to a substantial energy lowering at the HF level.
\insertnew{Evidently, despite the poor RHF reference, RMP2 recovers more correlation energy than cRMP2, perhaps because of overcorrelating small gap contributions.}

Lastly, we note that there is a significant energy difference between AP+UMP2 and cRMP2 similar to the \ce{O2} triplet-singlet gap result observed in Section \ref{sec:st12}.
However, this is mainly due to the qualitative difference between cR and U solutions. Spin-contamination often drives artifacts in the spin-density distribution which cannot be easily fixed by a posteriori spin projection methods. However, this can potentially be fixed by orbital optimizing in the presence of correlation as we shall see.

\begin{table}[h!]
  \centering
  \begin{tabular}{c|r|r}\Xhline{3\arrayrulewidth}
  Method & $\Delta E$ & $\langle \hat{S}^2\rangle_{M_S=0}$\\\Xhline{3\arrayrulewidth}
$\kappa$-ROOMP2 & 8.21 & 0.000\\\hline
$\kappa$-cROOMP2 & 10.17 & 0.000\\\hline
$\kappa$-UOOMP2 & 19.70 & 0.968\\\hline
AP+$\kappa$-UOOMP2 & 9.30 & 0.000\\\Xhline{3\arrayrulewidth}
U$\omega$B97X-V & 24.72 & 0.970\\\hline
AP+U$\omega$B97X-V & 15.80 & \\\hline
U$\omega$B97M-V & 20.42 & 0.943\\\hline
AP+U$\omega$B97M-V & 11.03 & \\\Xhline{3\arrayrulewidth}
  \end{tabular}
  \caption{
The activation energy $\Delta E$ (kcal/mol) of [Cys-H]$^{-}$+\ce{O2} from various OOMP2 and DFT methods.
The expectation values of $\langle \hat{S}^2\rangle$ for the $M_S=0$ state of the precursor are presented as well. 
Note that these values include correlation corrections to $\langle \hat{S}^2\rangle$ wherever appropriate.
  }
  \label{tab:oomp2singlet}
\end{table}

In Table \ref{tab:oomp2singlet}, we present the activation barrier obtained using various types of $\kappa$-OOMP2 methods and two popular, combinatorially optimized density functional theory (DFT) methods ($\omega$B97X-V\cite{Mardirossian2014} and $\omega$B97M-V\cite{Mardirossian2016}). 
%
%
First, we note that $\kappa$-cROOMP2 and AP+$\kappa$-UOOMP2 predict a barrier within 1 kcal/mol from each other.
This is because the $\kappa$-UOOMP2 $M_S=0$ state no longer has any artificial symmetry breaking and produces a solution with only essential spin-symmetry breaking.
$\kappa$-ROOMP2 is similar to $\kappa$-OOMP2 with cR or AP despite the lack of open-shell character in the wavefunction. 
The R to cR instability at the $\kappa$-OOMP2 level causes an energy lowering of only about 2 kcal/mol.
$\kappa$-UOOMP2 overestimates the gap by a factor of 2 compared to the corresponding AP results.

To see how well $\kappa$-cROOMP2 and AP+$\kappa$-UOOMP2 perform, we also compare this with two DFT methods. Without AP, both DFT methods with U orbitals predict the barrier too high. With AP, $\omega$B97X-V predicts a barrier of 15.80 kcal/mol while $\omega$B97M-V predicts a barrier of 11.03 kcal/mol. 
There is a quite significant functional dependence on the barrier height with the AP prescription \insertnew{(this may be related to the fact that $\langle \hat{S}^2\rangle$ cannot be rigorously evaluated: the expectation value of the KS determinant is used)}.
A barrier height of about 10 kcal/mol was obtained with $\kappa$-cROOMP2, AP+$\kappa$-cROOMP2, and AP+$\omega$B97M-V and an even higher height with AP+$\omega$B97X-V. All of these suggest that the reactivity of [Cys-H]$^-$ with \ce{O2} ($^1\Delta_g$) is moderate at room temperature.
In passing, we note that a higher level benchmark data would be desirable, using more sophisticated and computationally demanding methods such as equation of motion spin-flip coupled-cluster with singles and doubles (EOM-SF-CCSD).\cite{Krylov2001} This will be interesting to study in the future.

\subsection{Triplet-Singlet Gap of \ce{C30}}

\begin{figure}[h!]
\includegraphics[scale=0.05]{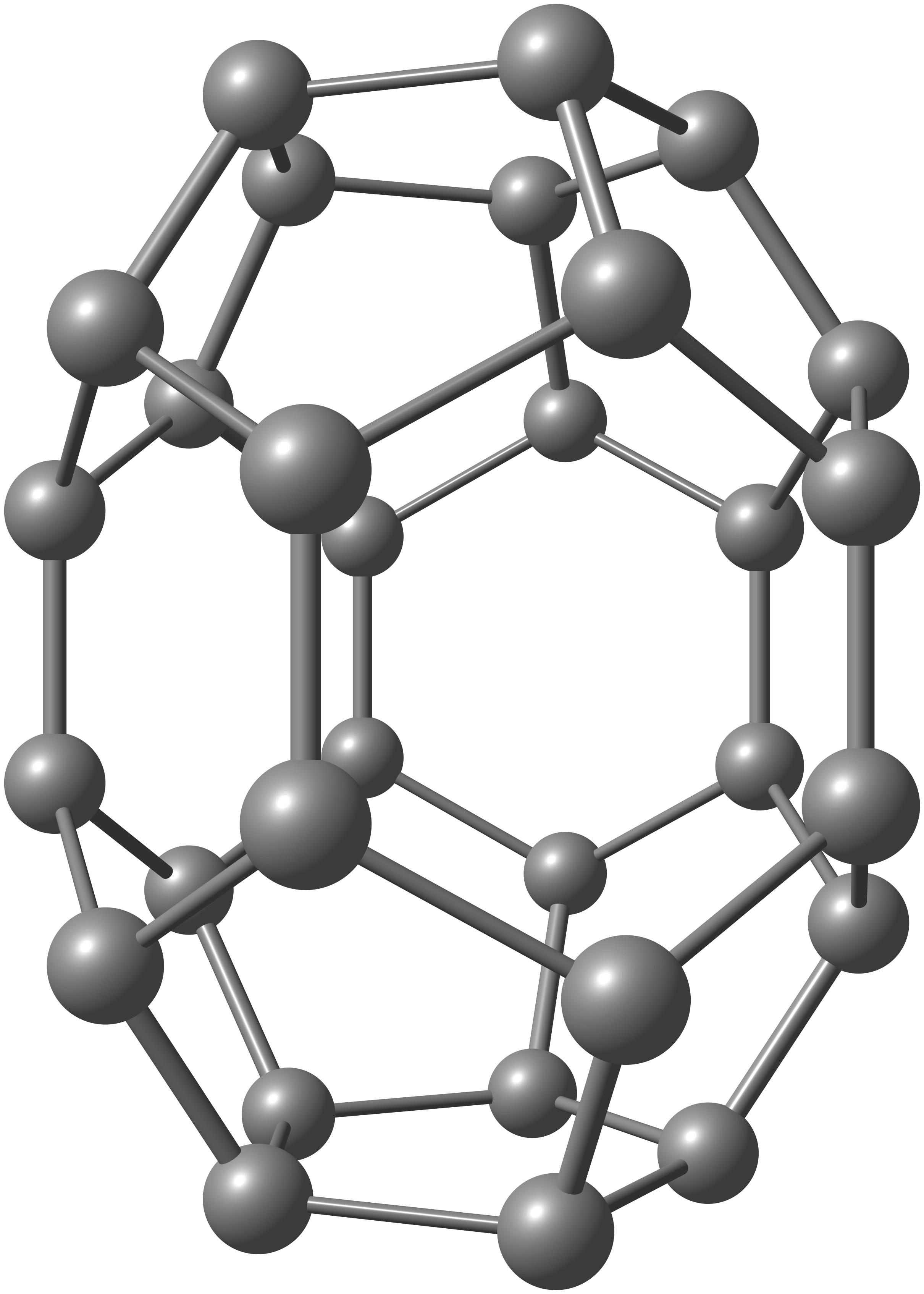}
\caption{
Molecular geometries of \ce{C30} ($D_\text{5h}$).
The Cartesian coordinates for this geometry used in this work are provided in the Supporting Information.
}
\label{fig:c30}
\end{figure}

Fullerenes are an interesting class of molecular clusters that are made solely of carbon atoms.
They all form intriguing cage structures and often are stable enough to be experimentally synthesized.
\ce{C30} is one of the smaller fullerenes and it has been quite challenging to isolate \ce{C30} experimentally due to its instability. 
It was pointed out in several experimental\cite{Kietzmann1998} and theoretical\cite{Zhang1992,Fan1995} studies that the highest symmetry structure, $D_\text{5h}$, is highly reactive.
This particular molecular geometry is presented in Figure \ref{fig:c30}.

The molecular geometries used in this work are optimized with unrestricted B97M-V\cite{mardirossian2015mapping} for each $M_S$ state, with $D_\text{5h}$ geometry within the cc-pVDZ basis set.\cite{Dunning1989} As this system is biradicaloid, the geometry of the $M_S=0$ state may require special care, but for simplicity we employed unrestricted calculations. The $\langle S^2 \rangle$ values of each state with this particular functional is 1.020 and 2.013, respectively. 
We provide the geometries of \ce{C30} used in this work in the Supporting Information. Details about the geometries will not alter the qualitative conclusion we are drawing in this section as long as the underlying point group symmetry is $D_\text{5h}$. 

Jim{\'e}nez-Hoyos and co-workers reported the existence of complex generalized HF (cGHF) solutions for \ce{C30} ($D_\text{5h}$) and concluded that \ce{C30} is a polyradicaloid based on the cGHF solution.\cite{Jimenez-Hoyos2014}
Due to its pronounced strong correlation, it is not surprising to observe symmetry breaking at the HF level.
However, one may wonder if breaking every symmetry from RHF to cGHF is {\it essential} since UHF is sufficient for most singlet biradicaloid systems.

We have developed a computational strategy which can identify {\it artificial} symmetry breaking at the HF level using $\kappa$-OOMP2 with cG orbitals.\cite{Lee2019}
We scan over a range of $\kappa$ values (i.e., the regularization strength) and compute the critical regularization strength, $\kappa_c$, to break/restore a given symmetry. Symmetry breaking with $\kappa_c$ $\in$ [0.0, 1.0] is categorized as {\it artificial} symmetry breaking, $\kappa_c$ $\in$ [1.0, 2.0] is {\it essential} symmetry breaking, and symmetry restoration for $\kappa_c$ $>$ 2.0 may be considered to be {\it artificial} restoration (i.e., too little symmetry breaking).
The symmetry landscape of \ce{C30} will help to identify the character of {\it essential} symmetry breaking in this system.

We obtained the symmetry breaking landscape of \ce{C30} within the 6-31G basis set\cite{Hehre1972} along with the cc-pVDZ auxiliary basis set.\cite{Weigend2002} The frozen core approximation was used for computational efficiency.
We focused on three symmetry breaking parameters: the spin expectation value $\langle \hat{S}^2 \rangle$, the non-collinearity order parameter $\mu$,\cite{Small2015} and the fundamental complexification measure $\xi$.\cite{Small2015a, Lee2019}

\begin{figure}[h!]
\includegraphics[scale=0.63]{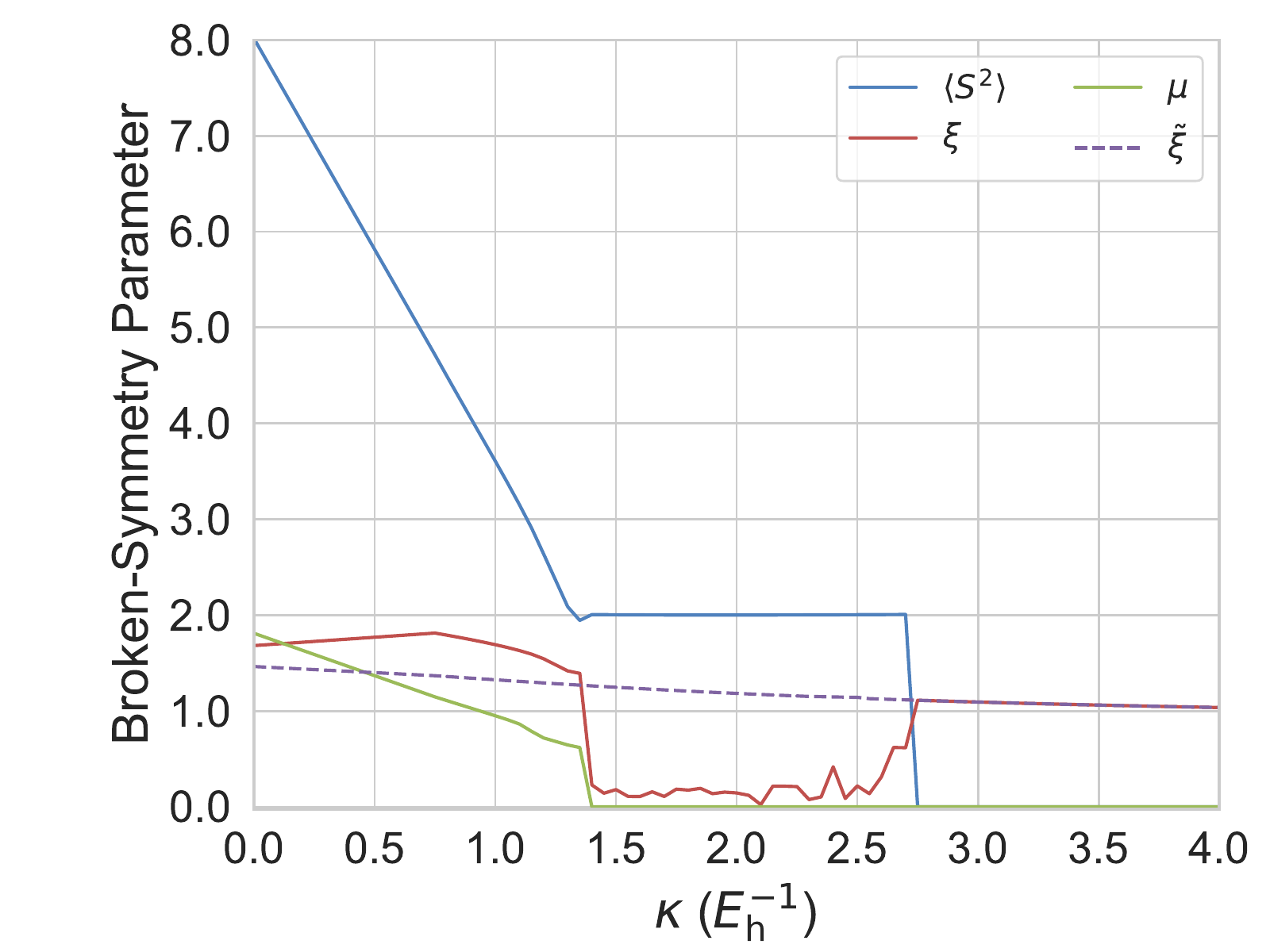}
\caption{
Measures of symmetry breaking ($\langle S^2\rangle$, $\xi$, and $\mu$) as a function of the regularization strength $\kappa$ for \ce{C30} (D$_\text{5h}$). 
$\tilde{\xi}$ is the complex broken-symmetry parameter of $\kappa$-cROOMP2.
These quantities characterize symmetry-breaking/restoration in $\kappa$-OOMP2.
}
\label{fig:c302}
\end{figure}

In Figure \ref{fig:c302}, we see that $\kappa_c = 1.40$ for $\mu$, $\kappa_c = 2.70$ for $\langle S^2\rangle$, and there is no obvious symmetry restoration for $\xi$. 
Compared to our previous work on characterizing other fullerenes such as \ce{C60} and \ce{C36}, this landscape is more complex than the well-known biradicaloid \ce{C36}. Between $\kappa = 1.40$ and $\kappa = 2.70$, cU solutions are found.
It is interesting that for $\kappa > 2.70$ cR solutions are most stable and \insertnew{there are no U or cU solutions}.
R solutions are commonly observed in the weak regularization regime, $\kappa > 2.0$, but cR solutions are quite unusual to observe.
It turns out that this complex symmetry breaking in $\kappa$-cROOMP2 exists for all $\kappa$ \revision{values as shown with the purple dashed line in Figure \ref{fig:c302}}.
With $\kappa$-cGOOMP2 with $\kappa<2.70$, $\kappa$-cROOMP2 solutions are higher in energy than other spin symmetry broken solutions.
This is why these solutions are only observed with very weak regularization in the landscape.
Based on these results, we conclude that the symmetry breaking of $\langle S^2\rangle$, $\xi$, and $\mu$ is {\it essential} and this molecule is strongly correlated.

\begin{table}[h!]
  \centering
  \begin{tabular}{c|r|r|r}\hline
Method & $\Delta E_\text{T-S} $ & $M_S=0$ & $M_S=1$ \\\Xhline{3\arrayrulewidth}
RHF & 10.94 & 0.00 & 2.00 \\ \hline
UHF & -13.07 & 7.18 & 8.15 \\ \hline
cRHF & 10.86 & 0.00 & 2.00 \\\Xhline{3\arrayrulewidth}
RMP2 & -25.62 & 0.00 & 2.00 \\ \hline
UMP2 & -8.39 & 6.34 & 7.33 \\ \hline
cRMP2 & -27.50 & 0.00 & 2.00 \\\Xhline{3\arrayrulewidth}
$\kappa$-ROOMP2 & 5.36  & 0.00 & 2.00 \\ \hline
$\kappa$-UOOMP2 & 1.84 & 1.02 & 2.00 \\ \hline
$\kappa$-cROOMP2 & 3.93 & 0.00 & 2.00 \\ \Xhline{3\arrayrulewidth}
AP+$\kappa$-UOOMP2 & 3.75 & & 2.00 \\
 \end{tabular}
  \caption{
The triplet-singlet gap $\Delta E_\text{T-S} (= E_S - E_T)$ (kcal/mol) of \ce{C30} from various methods.
The expectation values of $\langle \hat{S}^2\rangle$ for  $M_S=0$ and $M_S=1$ states are presented as well. 
Note that these values include correlation corrections to $\langle \hat{S}^2\rangle$.
  }
  \label{tab:c30}
\end{table}

We also computed the triplet-singlet gap of \ce{C30} using HF, MP2, and $\kappa$-OOMP2 methods with multiple types of orbitals.
At $\kappa=1.45$ with the cc-pVDZ basis set, we found a $\kappa$-cUOOMP2 solution (with $\langle S^2 \rangle = 2.0$) when started from a cGHF solution.
This triplet $\kappa$-cUOOMP2 solution was found to be almost exactly degenerate with a triplet $\kappa$-UOOMP2 solution.
Therefore, for the remaining discussion we employed $\kappa$-UOOMP2 for the triplet state.

The computed triplet-singlet gaps are presented in Table \ref{tab:c30}, which are obtained with the cc-pVTZ basis set.
Without correlation, RHF and cRHF predict very large gaps with a triplet ground state whereas UHF predicts a large gap with a singlet ground state.
The MP2 correction on top of these reference states all prefers the singlet state with a significant spin gap.
This is a qualitative failure of the MP2-level correlation treatment.

$\kappa$-OOMP2 provides a significant improvement over the MP2 results. 
\insertnew{$\kappa$-ROOMP2 predicts the sign of the gap correctly with a gap of 5.36 kcal/mol.}
$\kappa$-cROOMP2 yields a slightly smaller gap than $\kappa$-ROOMP2 and the energy lowering from complex polarization is only about 1.43 kcal/mol.
\insertnew{$\kappa$-UOOMP2 yields almost a perfect open-shell solution (i.e., $\langle S^2 \rangle \approx 1.0$), so AP+$\kappa$-UOOMP2 is effective for this system.}
\insertnew{AP+$\kappa$-UOOMP2  predicts more or less the same gap as $\kappa$-cROOMP2 and the difference between two is only 0.18 kcal/mol.}
In terms of the triplet-singlet gap, all of the $\kappa$-OOMP2 approaches predict the biradicaloid character of \ce{C30}.

Although the triplet-singlet gap from the R methods is similar to the cR methods, the use of R orbitals breaks the spatial symmetry (\ce{D_{5h}}) of \ce{C30}.
This is evident when looking at the Mulliken population of individual carbon atoms. To illustrate this, we present the Mulliken population of the five carbon atoms in the top pentagon of \ce{C30} in Figure \ref{fig:c30}. Obviously, they are all equivalent due to the \ce{D_{5h}} symmetry, but using restricted or unrestricted orbitals breaks this symmetry as shown in Table \ref{tab:charge}. Thus geometry optimization with other methods than cR methods will likely break this spatial symmetry. This is not because of the Jahn-Teller distortion but because of the {\it artificial} spatial symmetry breaking present at the electronic level.
\insertnew{In passing we mention than orbital-optimizing the spin-projected energy in Eq. \eqref{eq:Eproj} could potentially yield qualitatively better density than $\kappa$-UOOMP2.\cite{Mak2011}}

\begin{table}[h!]
  \centering
  \begin{tabular}{r|r|r|r|r|r}\hline
RHF & UHF & cRHF & $\kappa$-ROOMP2 & $\kappa$-UOOMP2 &$\kappa$-cROOMP2\\ \Xhline{3\arrayrulewidth}
0.0083 & 0.0001 & 0.0179 & -0.0027 & 0.0111 & 0.0099\\ \hline
0.0105 & 0.0001 & 0.0179 & 0.0067 & 0.0109 & 0.0099\\ \hline
0.0290 & -0.0034 & 0.0179 & 0.0187 & 0.0093 & 0.0099\\ \hline
0.0290 & 0.0056 & 0.0179 & 0.0187 & 0.0120 & 0.0099\\ \hline
0.0105 & -0.0034 & 0.0179 & 0.0067 & 0.0092 & 0.0099
 \end{tabular}
  \caption{
Mulliken population of the five carbon atoms in the top pentagon in \ce{C30} shown in Figure \ref{fig:c30}.
  }
  \label{tab:charge}
\end{table}

In summary, in this example we showed that $\kappa$-cROOMP2 is better suited than AP+$\kappa$-UOOMP2 in the presence of high point group symmetry such as D$_\text{5h}$.
Although they both yield similar energies, the underlying wavefunction breaks spatial symmetry if not treated with cR orbitals.
This highlights the unique utility of electronic structure methods with cR orbitals whenever complex polarization is {\it essential}.

\subsection{Stable Organic Triplet Biradical}
\begin{figure}[h!]
\includegraphics[scale=0.075]{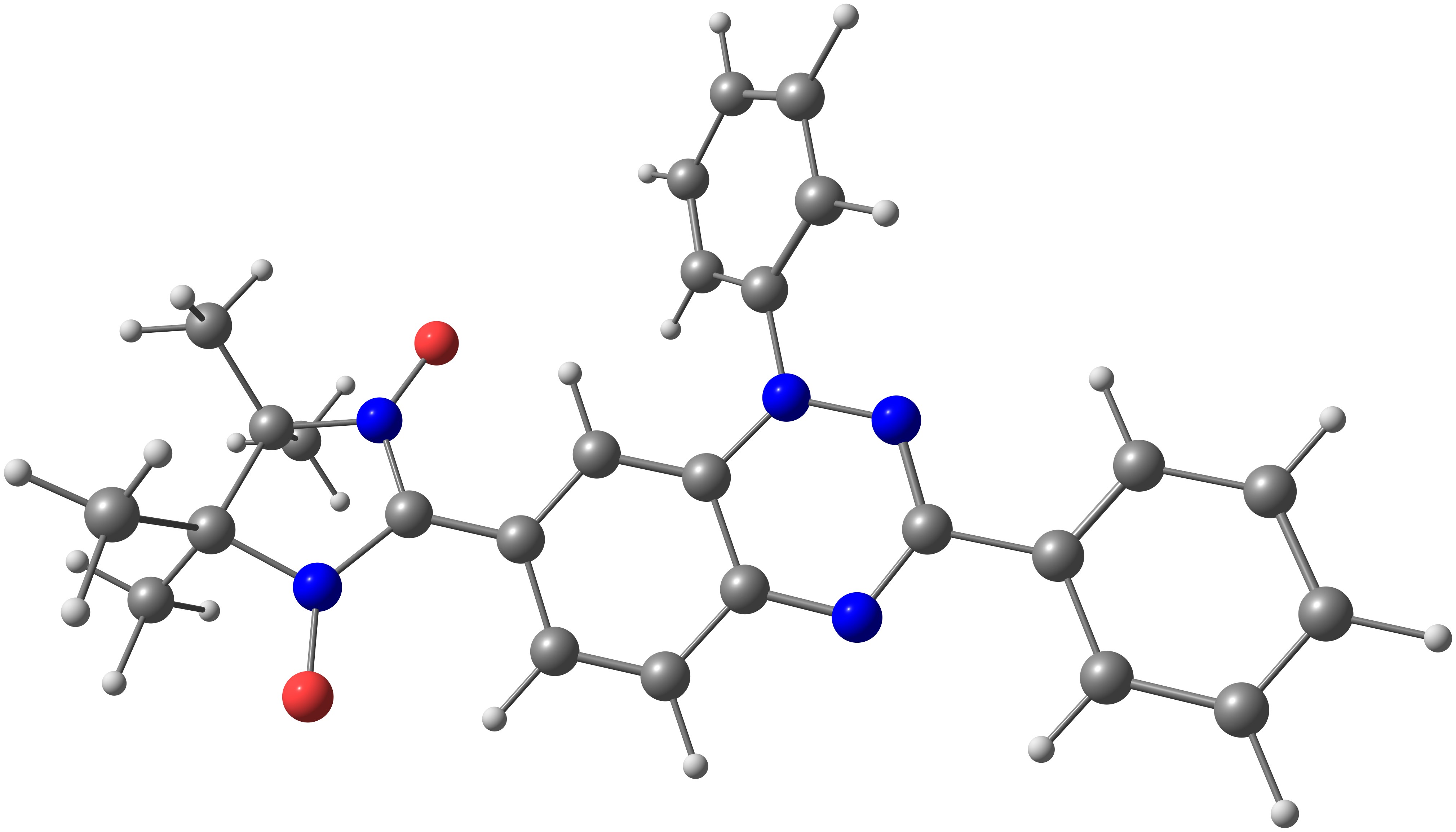}
\caption{
Molecular geometries of the organic biradical studied here.
The Cartesian coordinates for this geometry used in this work are taken from ref. \citenum{Gallagher2016}.
}
\label{fig:bi}
\end{figure}
Although organic triplet biradicals are very rare to isolate due to their normally high reactivity, there have been some reports of synthesizing stable ones.\cite{Rajca1994,Abe2013} 
\insertnew{Indeed, many stable singlet biradicaloids are stable because of some closed shell character.\cite{Jung2003}
Since triplet biradicals lack in any closed shell character, it is difficult to observe stable ones.}
Gallagher and co-workers synthesized an organic biradical with a triplet ground state.\cite{Gallagher2016} This biradical has quite robust stability compared to usual biradicals and survives at 140 $^\circ$C without significant decomposition. Experimentally the triplet-singlet gap of this molecule was measured to be about 0.5 kcal/mol. However, such a small gap allows for a thermal mixture of singlet and triplet states \insertnew{as temperature is raised to ambient conditions and above.}

Gallagher and co-workers suggested a modification to this synthesized complex and hypothesized a triplet ground state, aiming for a larger triplet-singlet gap than 0.5 kcal/mol.\cite{Gallagher2016} The structure of this proposed molecule is presented in Figure \ref{fig:bi}. They supported their claim using UB3LYP/6-31G(d,p) calculations which yielded a gap of 3.5 kcal/mol for this newly suggested complex.
Our goal is to confirm whether this hypothesis is correct using $\kappa$-cROOMP2 and/or AP+$\kappa$-UOOMP2.
We studied this system within the cc-pVDZ basis set\cite{Dunning1989} and its auxiliary basis set\cite{Weigend2002} with the frozen-core approximation and the geometries were taken from ref. \citenum{Gallagher2016} which were optimized at the UB3LYP/6-31G(d,p) level.
\insertnew{
This proposed system was recently synthesized and characterized with $\Delta E_{TS} \ge 1.7$ kcal/mol.\cite{Gallagher2019}
}

Unlike other examples presented above, there are no obvious symmetry constraints that give rise to a R to cR instability in this system.
This is why it is interesting that there is a R to cR instability at the HF level (see Table \ref{tab:birad}).
However, this complex polarization turns out to be {\it artificial} and $\kappa$-OOMP2 with $\kappa=1.45$ yields only a restricted solution.
Therefore, in this case $\kappa$-cROOMP2 is not applicable whereas AP+$\kappa$-UOOMP2 is well-suited.

\begin{table}[h!]
  \centering
  \begin{tabular}{c|r|r|r}\hline
Method & $\Delta E_\text{T-S} $ & $M_S=0$ & $M_S=1$ \\\Xhline{3\arrayrulewidth}
RHF & 63.56 & 0.00 & 2.00\\\hline
UHF & 12.44 & 6.25 & 7.69\\\hline
cRHF & 63.49 & 0.00 & 2.00\\\Xhline{3\arrayrulewidth}
RMP2 & 29.88 & 0.00 & 2.00\\\hline
UMP2 & -8.64 & 2.56 & 3.86\\\hline
cRMP2 & 48.97 & 0.00 & 2.00\\\Xhline{3\arrayrulewidth}
$\kappa$-UOOMP2 & 1.48 & 1.02 & 2.03\\\hline
AP+$\kappa$-UOOMP2 & 2.97 &  & 2.00\\\Xhline{3\arrayrulewidth}
$\kappa$-cR/ROOMP2 & 36.23 & 0.00 & 2.00\\\hline
 \end{tabular}
  \caption{
The triplet-singlet gap $\Delta E_\text{T-S} (= E_S - E_T)$ (kcal/mol) of the biradical system in Figure \ref{fig:bi} from various methods.
The expectation values of $\langle \hat{S}^2\rangle$ for  $M_S=0$ and $M_S=1$ states are presented as well. 
Note that these values include correlation corrections to $\langle \hat{S}^2\rangle$.
  }
  \label{tab:birad}
\end{table}

In Table \ref{tab:birad}, the triplet-singlet gap of this system is presented.
At the HF level, none of the orbital types predict small enough gaps to be considered to be a biradical.
RHF and cRHF states are nearly degenerate and thus the complex polarization is not as strong as other examples presented before.
UHF exhibits striking spin-symmetry breaking and predicts a much smaller spin gap than RHF and cRHF.

The MP2 treatment on top of these reference HF determinants does not improve these poor energetics.
There is about a 20 kcal/mol energy difference between RMP2 and cRMP2 and RMP2 is lower in energy than cRMP2.
This may indicate {\it artificial} complex polarization which indeed turns out to be the case in this system (vide infra).
UMP2 removes a large portion of the spin contamination present at the HF level, but it still is heavily spin-contaminated.
As a result, it predicts the sign of the gap incorrectly.

$\kappa$-UOOMP2 predicts a reasonably small gap with satisfying spin contamination for the singlet state ($\langle S^2 \rangle \approx 1.0$) and almost no
spin contamination for the triplet state. With the AP scheme, the gap is predicted to be 2.97 kcal/mol. This supports the original hypothesis\cite{Gallagher2016} made by experimentalists that this system has a gap larger than 0.5 kcal/mol, with a triplet ground state. This is also in agreement with the recent experiment which studied this system.\cite{Gallagher2019}
Lastly, we note that $\kappa$-cROOMP2 collapses to a real, restricted solution and yields a substantially larger gap (36.23 kcal/mol), \insertnew{because this method does not adequately describe the strongly correlated singlet}.

In summary, in this example, AP+$\kappa$-UOOMP2 successfully describes the biradicaloid character of the singlet state in the molecule whereas 
$\kappa$-cROOMP2 cannot describe such character because there is no cR solution at the $\kappa$-OOMP2 level.
\section{Conclusions}

In this work, we examined two single-reference approaches based on regularized orbital-optimized M{\o}ller-Plesset perturbation theory ($\kappa$-OOMP2) that exploit {\it essential} symmetry breaking to describe singlet biradicaloids.
Combined with Yamaguchi's approximate projection (AP), unrestricted $\kappa$-OOMP2 ($\kappa$-UOOMP2) offers a way to access almost spin-pure singlet energies.
Alternatively, complex, restricted $\kappa$-OOMP2 ($\kappa$-cROOMP2) can describe biradicaloid character if there is complex polarization.
We compared these two methods over a variety of systems: a total of 12 triplet-singlet gaps in the TS12 set, the barrier height of a reaction between a cysteine ion and a singlet oxygen molecule, the \ce{C30} (D$_\text{5h}$) fullerene, and lastly an organic biradical with a triplet ground state. We summarize the major conclusions from these numerical experiments as follows:
\begin{enumerate}
\item Without orbital optimization at the MP2 level, Hartree-Fock (HF) orbitals tend to exhibit {\it artificial} symmetry breaking in singlet biradicaloids. In the case of cRHF, this is sometimes reflected in spurious charge distribution of molecules whereas it often manifests as heavy spin contamination (and commonly also spurious charge distribution) in UHF.
In such cases, we recommend $\kappa$-OOMP2 which is an electronic structure tool that removes most {\it artificial} symmetry breaking and yields orbitals with only {\it essential} symmetry breaking.
\item $\kappa$-cROOMP2 is recommended whenever there is {\it essential} complex polarization. This is due to the fact that $\kappa$-UOOMP2 manifests not only spin-symmetry breaking but also spatial symmetry breaking which cannot be purified with the AP scheme.
\item When there is no {\it essential} complex polarization but only {\it essential} spin polarization, AP+$\kappa$-UOOMP2 is recommended. cR solutions are rare in nature and it is difficult to observe them with systems without point group symmetry. Therefore, the applicability of AP+$\kappa$-UOOMP2 is broader than that of $\kappa$-cROOMP2.
\end{enumerate}
Strong correlation is a difficult problem to solve and there is no universal approach to it other than brute-force approaches such as complete active space methods.\cite{Szalay2012} However, at least for two-electron strong correlation problems studied here, either $\kappa$-cROOMP2 or AP+$\kappa$-UOOMP2 can be a single-reference electronic structure method that correctly describes strong correlation character. It will be interesting to apply these tools to a broader range of chemical systems along with more developments on their response theory such as excited states and analytic nuclear gradients in the future.
\insertnew{The presented approaches, which use cR orbitals or AP, can be extended to higher order single-reference correlation methods such as coupled-cluster with singles and doubles (CCSD) and third-order M{\o}ller-Plesset perturbation theory (MP3).}
\section{Supplementary Material}
The supplemental material of this work is available online which discusses the frozen core and frozen virtual approximation in $\kappa$-OOMP2, the theoretical reference data of TS12 set and the Cartesian coordinate of \ce{C30}.
\section{Acknowledgement}
This work was supported by a subcontract from MURI grant W911 NF-14-1-0359. J. L. thanks Soojin Lee for consistent encouragement.

\section{Supplementary Material}
The supplemental material of this work is available online which includes the proof of Eq. \eqref{eq:zeroquad}, Eq. \eqref{eq:T3} in terms of computable quantities and the CCVB+i3 Jacobian, Lagrangian and associated derivatives for optimization.
\bibliography{croomp2_ms}
\bibliographystyle{achemso}
\end{document}